\documentclass[prb]{revtex4}
\usepackage{graphicx}
\usepackage{CJK}
\usepackage{graphicx}

\begin{document}
	\begin{CJK*}{GBK}{song}
		\title{Anti-Chaos Control via Nonlinear Schr\"{o}dinger Equations for the secured optical communication }
		\author{Zhenyu Tang} \email{zhenyutang2011@gmail.com} \author{Hui-Ling Zhen}
		
		\begin{abstract}
			Coupled nonlinear Schr\"{o}dinger equations, governing the propagation of envelopes of
			electromagnetic waves in birefringent optical fibers, are studied in this paper for
			their potential applications in the secured optical communication. Periodicity and
			integrability of the CNLS equations are obtained via the phase-plane analysis. With
			the time-delay and perturbations introduced, CNLS equations are chaotified and a
			chaotic system is proposed. Numerical and analytical methods are conducted on such
			system: (I) Phase projections are given and the final chaotic states can be observed.
			(II) Power spectra and the largest Lyapunov exponents are calculated to corroborate
			that those motions are indeed chaotic. 
			
		\end{abstract}
		
		
		\maketitle
		
		
		Interest in the optical solitons has grown for their potential applications in
		the telecommunications and ultrafast signal routing systems, especially the vector
		ones~\cite{beijingsoliton1,beijingsoliton2}. The coupled nonlinear Schr\"{o}dinger
		(CNLS) equations, which can be used to govern the propagation of envelopes of
		electromagnetic waves in birefringent optical fibers, read as~\cite{source1,source2},
		\begin{eqnarray}\label{fangcheng1}
		&& \hspace{-1.5cm}  i u_x+u_{tt}+2 \kappa (|u|^2+|v|^2) u=0,\\
		&& \hspace{-1.5cm}  i v_x+v_{tt}+2 \kappa (|u|^2+|v|^2) v=0,
		\end{eqnarray}
		where $u$ and $v$, two complex functions about $x$ and $t$, are the normalized envelopes
		of the optical pulses along the two circularly polarized modes of a birefringent optical
		fiber, $x$ represents the normalized distance along the direction of propagation, $t$
		refers to the retarded time, and $\kappa$ gives the strength of the nonlinearity~\cite{source1,source2}.
		Some studies on Eqs.~(\ref{fangcheng1}) have been investigated in the early literatures,
		e.g., the bright soliton solutions~\cite{done1}, dark soliton solutions~\cite{done2}
		and the effects of noise on the solitons~\cite{done3}.
		
		Chaotic dynamics, owing to its noise-like broadband power spectra, is a good candidate
		to fight narrow-band effects, such as the frequency-selective fading or narrow-band
		disturbances in the communication systems~\cite{secure3,secure4}. Thus, for the secured
		optical communication, chaotic signals have received increasing attention because of
		their dependence on the initial condition, which makes it difficult to guess the
		structure of the generator and to predict the signal over a longer time interval~\cite{secure1,secure2}.
		Therefore, as opposed to controlling or eliminating chaos in dynamical systems~\cite{control},
		creating chaos from a non-chaotic system attracts some interests for the secured
		optical communication and information security~\cite{chaotification1}. People have
		known that a system with time-delay is inherently infinite dimensional, so it can
		produce complicated dynamics such as bifurcation and chaos, even a first-order
		system~\cite{timedelay1,timedelay2}. So the time-delay feedback method has been
		thought as a straightforward one to chaotify a non-chaotic system~\cite{chaotification2}.
		
		Early literatures have investigated that the dark solitons array can be used in
		the secured optical communication~\cite{dark1,dark2}, and the results are claimed
		to benefit the study on soliton equations in such field~\cite{dark1}. However, to
		our knowledge, little work has been done on Eqs.~(\ref{fangcheng1}) for their
		potential applications in the secured optical communication. In this paper, as
		an interest in chaos, analytical and numerical studies will be conducted on Eqs.~(\ref{fangcheng1})
		to reveal their potential applications in this field.
		
		Setting $u(x,t)=\varphi(\xi)e^{i \vartheta_1}$, $v(x,t)=\psi(\xi) e^{i \vartheta_2}$
		with $\xi=a_1 x-b_1 t$, $\vartheta_1=a_2 x-b_2 t$ and $\vartheta_2=a_3 x-b_3 t$, and
		substituting them into Eqs.~(\ref{fangcheng1}), we have
		\begin{eqnarray}\label{xitongqian}
		&& \hspace{-1.5cm}  \varphi_{\xi\xi}-o_1 \varphi+o_2 \varphi^3+o_2 \psi^2 \varphi=0,\\
		&& \hspace{-1.5cm}  \psi_{\xi\xi}-o_3 \psi+o_4 \psi^3+o_4 \varphi^2 \psi=0,
		\end{eqnarray}
		where
		\begin{eqnarray}
		&& \hspace{-1.5cm} o_1=\frac{a_2+b_2^2}{b_1^2},\ \ o_2=\frac{2 \kappa}{b_1^2},
		\ \ o_3=\frac{a_3+b_3^2}{b_1^2},\ \ o_4=\frac{2 \kappa}{b_1^2},\nonumber
		\end{eqnarray}
		with $a_j$'s and $b_j$'s ($j=1,2,3$) all being real constants.
		
		To investigate the dynamical characteristics of Eqs.~(\ref{fangcheng1}), we rewrite
		Eqs.~(\ref{xitongqian}) in the form of a four-dimensional planar dynamic system as follows
		($X_1 \equiv \varphi$, $X_2 \equiv \psi$, $Y_1 \equiv \varphi_\xi$, $Y_2 \equiv \psi_\xi$):
		\begin{equation}\label{xitong}
		\hspace{-1.7cm} \left\{
		\begin{array}{l}
		\vspace{1.5mm} X_{1,\xi}=Y_1, \ \ \ X_{2,\xi}=Y_2, \\
		\vspace{1.5mm} Y_{1,\xi}=o_1 X_1-o_2 X_1^3-o_2 X_2^2 X_1, \ \ \ Y_{2,\xi}=o_3 X_2-o_4 X_2^3-o_4 X_1^2 X_2.
		\end{array}
		\right.
		\end{equation}
		Phase projections for System~(\ref{xitong}) are shown in Figs.~1, and power spectra for
		the solutions of System~(\ref{xitong}) are calculated in Figs.~2.
		
		\begin{figure}
			\includegraphics[width=3in]{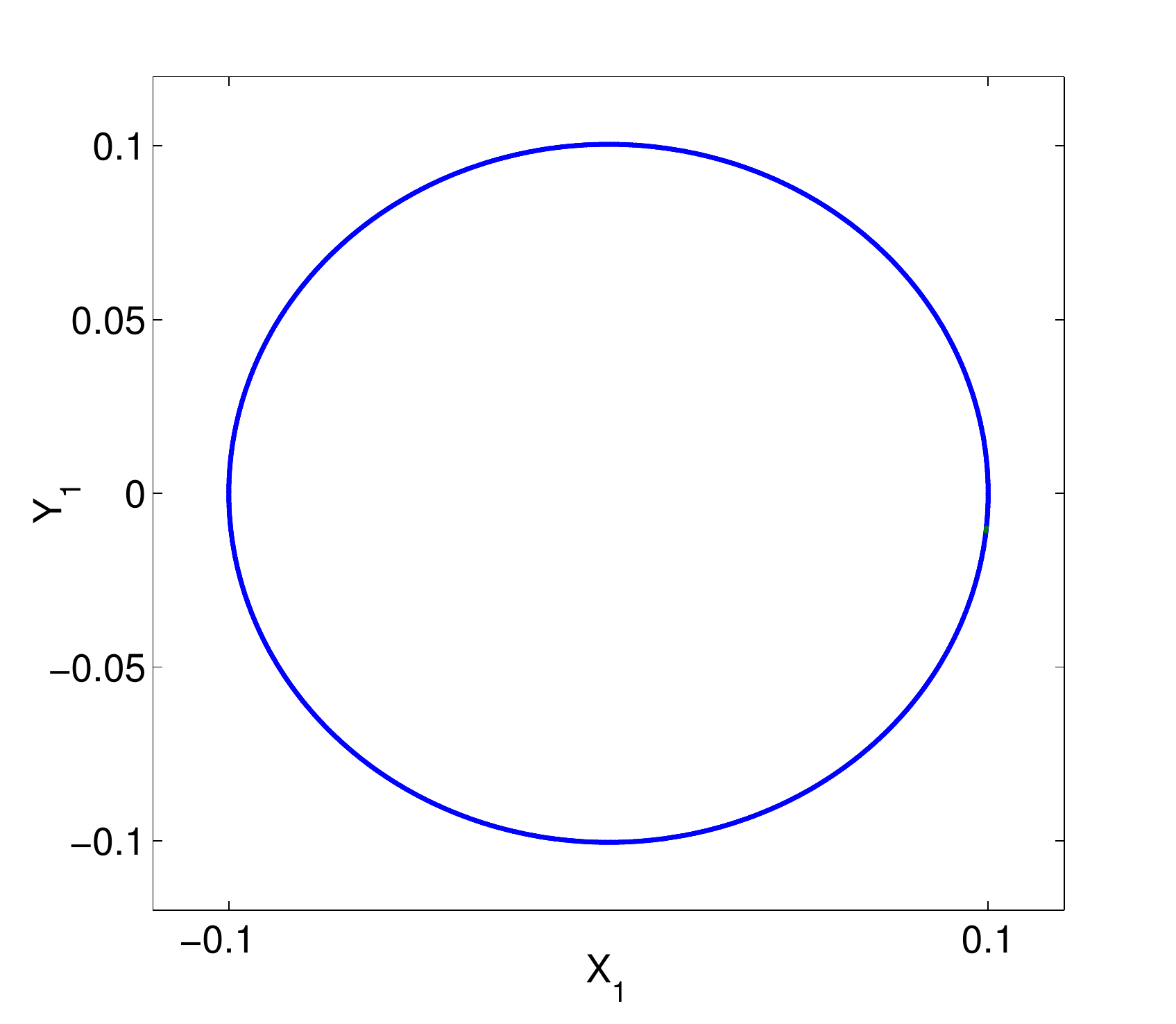}
			\caption{\label{pi}(a) Phase projections of System~(\ref{xitong}) with $o_1$=1, $o_2$=1.5, $o_3$=0.5, $o_4$=0.8, $Y_2$=1.}
		\end{figure}
		
		\begin{figure}
			\includegraphics[width=3in]{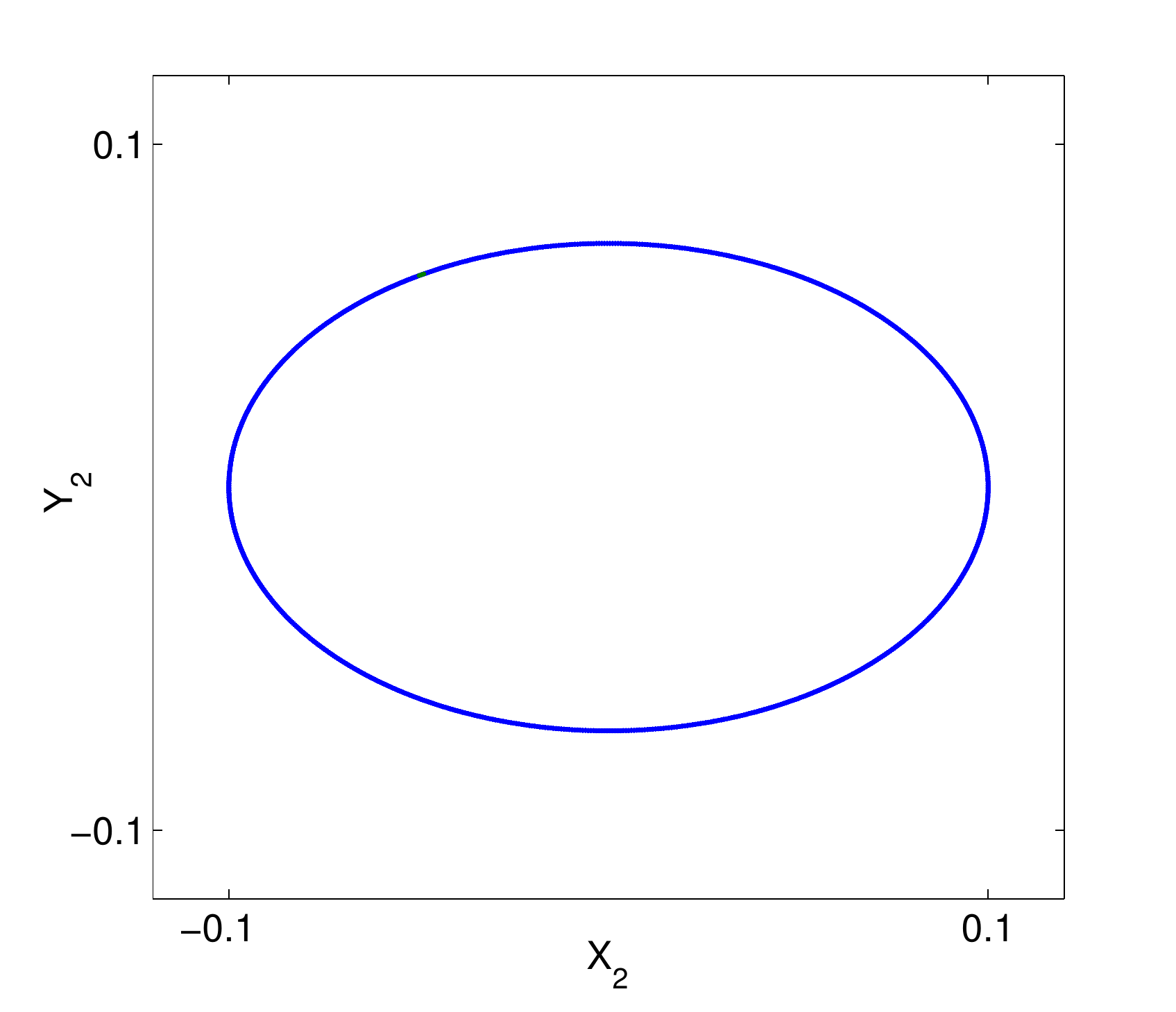}
			\caption{\label{pi}(b) The same as 1(a) but $Y_1$=1.}
		\end{figure}
		
		\begin{figure}
			\includegraphics[width=3in]{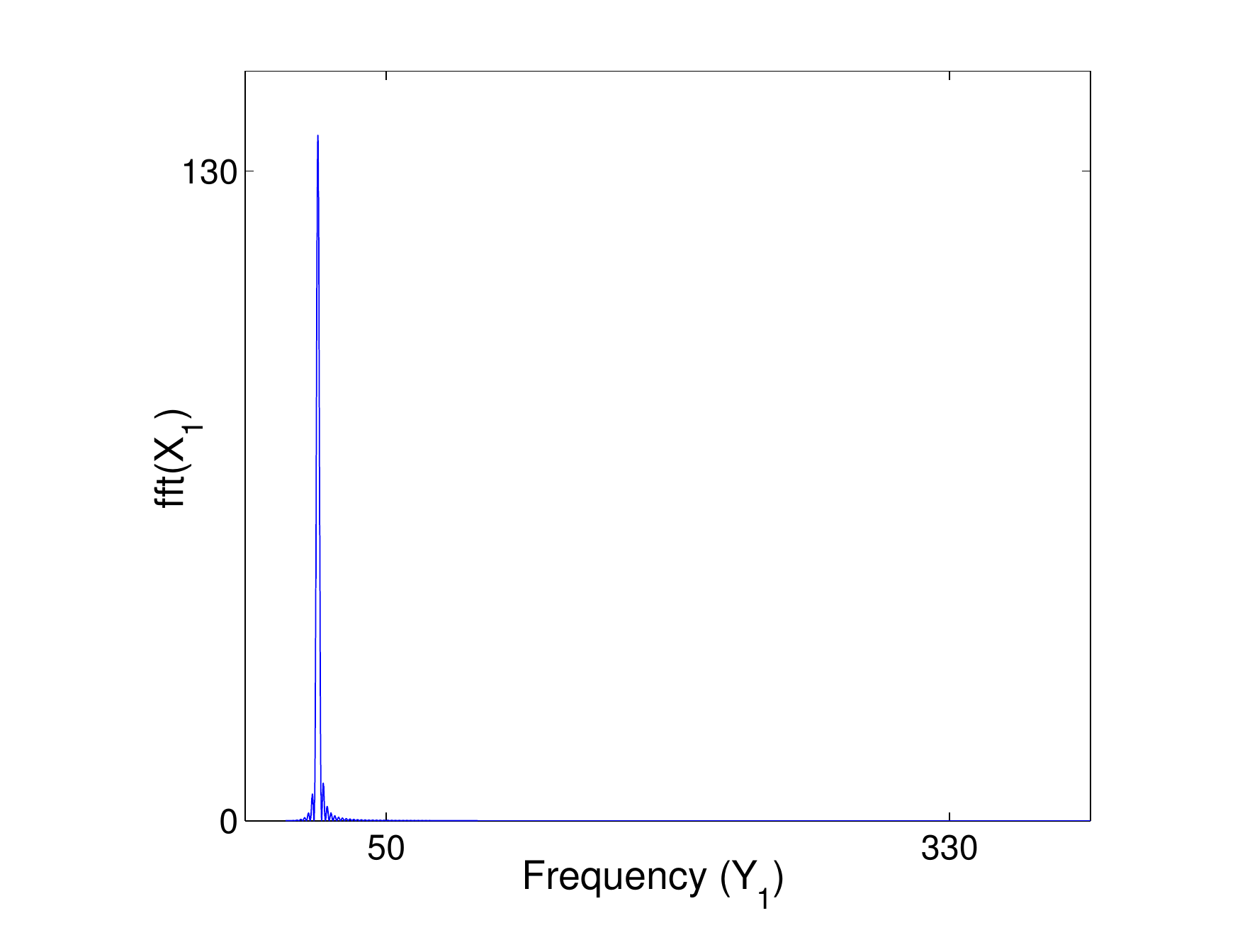}
			\caption{\label{pi}(a) Power spectra for the solutions of System~(\ref{xitong}) which correspond with 1(a).}
		\end{figure}
		
		\begin{figure}
			\includegraphics[width=3in]{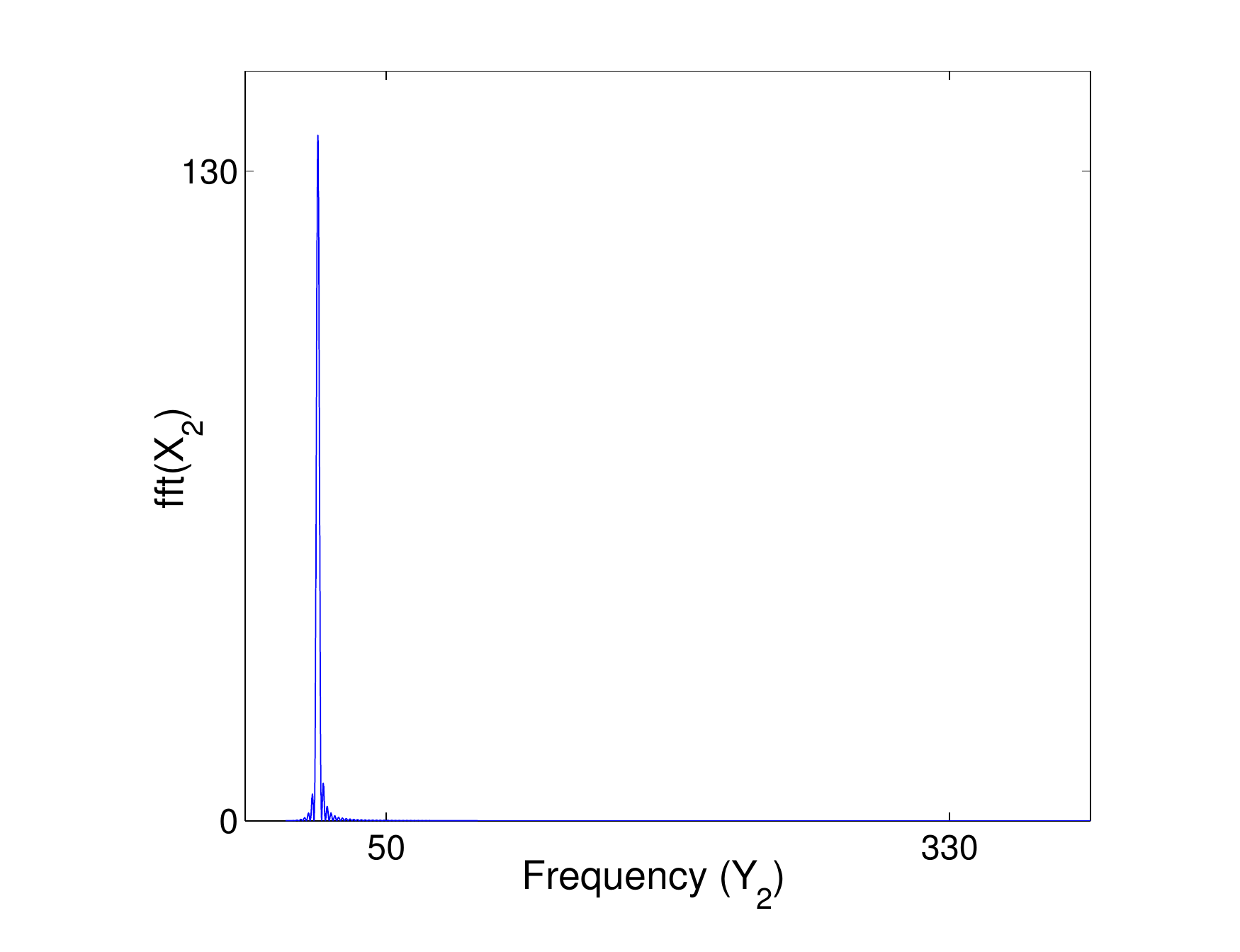}
			\caption{\label{pi}(b) Power spectra for the solutions of System~(\ref{xitong}) which correspond with 1(b).}
		\end{figure}
		
		From Figs.~1, we can see the closed curves, which can be used to represent the phase
		projections of System~(\ref{xitong}). Based on the power spectra in Figs.~2, periodicity
		of Eqs.~(\ref{fangcheng1}) is verified owing to the single frequency in Figs.~2. Thus,
		owing to the conclusions in Refs.~\cite{dynamic1,dynamic2}, we know that System~(\ref{xitong})
		is integrable, and Eqs.~(\ref{fangcheng1}) do not admit any chaotic motions. Hereby,
		$fft(X_1)$ and $fft(X_2)$ represent the fast Fourier transform (FFT) of $X_1$ and
		$X_2$, respectively~\cite{fft}.
		
		System~(\ref{xitong}) can be rewritten as
		\begin{eqnarray}\label{raodongqian1}
		\left(\begin{array}{c}
		X_1 \\
		X_2 \\
		Y_1 \\
		Y_2
		\end{array} \right)_\xi =\left(\begin{array}{c}
		Y_1 \\
		Y_2 \\
		o_1 X_1-o_2 X_1^3-o_2 X_2^2 X_1 \\
		o_3 X_2-o_4 X_2^3-o_4 X_1^2 X_2
		\end{array} \right),
		\end{eqnarray}
		with $\mathbf{x_1}=(0,0,0,0)^{'}$, $\mathbf{x_2}=\left( \pm \sqrt{\frac{o_1}{o_2}},0,0,0 \right)^{'}$ and
		$\mathbf{x_3}=\left( 0,\pm \sqrt{\frac{o_3}{o_4}},0,0 \right)^{'}$ being its equilibrium points,
		where $'$ denotes the vector transpose. Without loss of generality, we choose $o_1$
		and $o_3$ as the control parameters.
		
		According to the time-delay feedback method~\cite{chaotification1,chaotification2}, chaotifying
		System~(\ref{xitong}) is equivalent to to construct a single-input single-output
		system as follows:
		\begin{equation}\label{moxing1}
		\hspace{-1.7cm} \left\{
		\begin{array}{l}
		\vspace{1.5mm} \mathbf{x}_\xi=\mathbf{f}(\mathbf{x})+\mathbf{g}(\mathbf{x})\delta(\xi), \\
		\vspace{1.5mm} y=h(\mathbf{x}),
		\end{array}
		\right.
		\end{equation}
		where $\mathbf{x}$ and $y$ label the input and output, respectively,
		$\textbf{x}_\xi=d \textbf{x}/d \xi$, $\mathbf{f}(\mathbf{x})$ and $\mathbf{g}(\mathbf{x})$
		are both real vector functions, $\delta(\xi)$ corresponds to a system parameter
		perturbation or an exogenous control input, and $h(\mathbf{x})$ is a smooth real
		function and refers to the output~\cite{chaotification1,chaotification2}.
		Hereby, in the case of System~(\ref{raodongqian1}), $\mathbf{x}=(X_1,X_2,Y_1,Y_2)^{'}$,
		while $\mathbf{f}(\mathbf{x})$ and $\mathbf{g}(\mathbf{x})$ can be given as
		\begin{eqnarray}\label{fg}
		&& \hspace{-1.5cm} \mathbf{f}(\mathbf{x})=\left(\begin{array}{c}
		Y_1 \\
		Y_2 \\
		o_1 X_1-o_2 X_1^3-o_2 X_2^2 X_1 \\
		o_3 X_2-o_4 X_2^3-o_4 X_1^2 X_2
		\end{array} \right),\ \ \
		\mathbf{g}(\mathbf{x})=\left(\begin{array}{c}
		0 \\
		0 \\
		X_1 \\
		X_2
		\end{array} \right).
		\end{eqnarray}
		Then, System~(\ref{moxing1}) can be embodied as
		\begin{equation}\label{raodongqian2}
		\hspace{-1.7cm} \left\{
		\begin{array}{l}
		\vspace{1.5mm} \mathbf{x}_\xi=\left(\begin{array}{c}
		Y_1 \\
		Y_2 \\
		o_1 X_1-o_2 X_1^3-o_2 X_2^2 X_1 \\
		o_3 X_2-o_4 X_2^3-o_4 X_1^2 X_2
		\end{array} \right)+\left(\begin{array}{c}
		0 \\
		0 \\
		X_1 \\
		X_2
		\end{array} \right) \delta(\xi), \\
		\vspace{1.5mm} y=h(\mathbf{x}),
		\end{array}
		\right.
		\end{equation}
		where $h(\mathbf{x})$ and $\delta(\xi)$ are to be determined.
		
		Based on Expression~(\ref{fg}), we have
		\begin{eqnarray}\label{ad}
		&& \hspace{-1.5cm} \mathbf{ad_f g}(\mathbf{x})=\left(\begin{array}{c}
		-X_1 \\
		-X_2 \\
		Y_1 \\
		Y_2
		\end{array} \right),\ \ \
		\mathbf{ad_f}^2 \mathbf{g}(\mathbf{x})=\left(\begin{array}{c}
		-2 Y_1 \\
		-2 Y_2 \\
		2 o_1 X_1-4 o_2 X_1^3+(2 o_1-2 o_2) X_1 X_2^2 \\
		2 o_3 X_2-4 o_4 X_2^3+(2 o_3-2 o_4) X_2 X_1^2
		\end{array} \right),  \\
		&& \hspace{-1.5cm} \mathbf{ad_f}^3 \mathbf{g}(\mathbf{x})=\left(\begin{array}{c}
		-4 o_1 X_1+6 o_2 X_1^3+(4 o_2-2 o_1) X_1 X_2^2 \\
		-4 o_3 X_x+6 o_4 X_2^3+(4 o_4-2 o_3) X_2 X_1^2 \\
		Y_1 [4 o_1-18 o_2 X_1^2+(2 o_1-4 o_2)X_2^2] +(4 o_1-8 o_2) X_1 X_2 Y_2 \\
		Y_1 [4 o_3-18 o_4 X_2^2+(2 o_3-4 o_4)X_1^2] +(4 o_3-8 o_4) X_1 X_2 Y_1
		\end{array} \right),
		\end{eqnarray}
		where $\textbf{ad}_\textbf{f} \textbf{g}(\textbf{x})$ refers to the Lie bracket~\cite{chaotification1,chaotification2}
		of the two smooth vector functions $\mathbf{f}(\mathbf{x})$ and $\mathbf{g}(\mathbf{x})$.
		Note that the relative degree of System~(\ref{raodongqian2}) is four, i.e., the dimension
		of $\mathbf{x}$, and the definition of ``relative degree" can be seen in
		Refs.~\cite{chaotification1,chaotification2}.
		
		Via the conclusions in Refs.~\cite{chaotification1,chaotification2}, $h(\mathbf{x})$
		should satisfy
		\begin{eqnarray}
		&& \hspace{-1.5cm} \frac{\partial h(\mathbf{x})}{\partial \mathbf{x}}
		[\mathbf{g}(\mathbf{x}),\mathbf{ad_f g}(\mathbf{x}),\mathbf{ad_f}^2 \mathbf{g}(\mathbf{x})]=0. \nonumber
		\end{eqnarray}
		Based on some calculations, it means that $h(\mathbf{x})$ can be expressed as
		\begin{eqnarray}\label{hx}
		&& \hspace{-1.5cm} h(\mathbf{x})=X_2 Y_1-X_1 Y_2,
		\end{eqnarray}
		which gives rise to the expressions of $\delta(\xi)$ as follows:
		\begin{eqnarray}
		&& \hspace{-1.5cm} \delta(\xi)=\varsigma \sin [ \sigma ( X_2(\xi-\tau) Y_1(\xi-\tau)-X_1(\xi-\tau) Y_2(\xi-\tau) ) ],
		\end{eqnarray}
		where $\varsigma$ and $\sigma$ are both the real constants, $\tau$ refers to the time-delay, and
		$\xi$ is given in Sec.~2.
		
		Therefore, based on the chaotification of Eqs.~(\ref{fangcheng1}), we can propose a chaotic
		system as follows:
		\begin{eqnarray}\label{raodonghou}
		\left(\begin{array}{c}
		X_1 \\
		X_2 \\
		Y_1 \\
		Y_2
		\end{array} \right)_\xi =\left(\begin{array}{c}
		Y_1 \\
		Y_2 \\
		(o_1+\delta) X_1-o_2 X_1^3-o_2 X_2^2 X_1 \\
		(o_3+\delta) X_2-o_4 X_2^3-o_4 X_1^2 X_2
		\end{array} \right) ,
		\end{eqnarray}
		where $\delta=\delta(\xi)=\varsigma \sin [ \sigma ( X_2(\xi-\tau) Y_1(\xi-\tau)-X_1(\xi-\tau) Y_2(\xi-\tau) ) ]$
		can be used as the perturbations of the control parameters $o_1$ and $o_3$.
		
		\begin{figure}
			\includegraphics[width=3in]{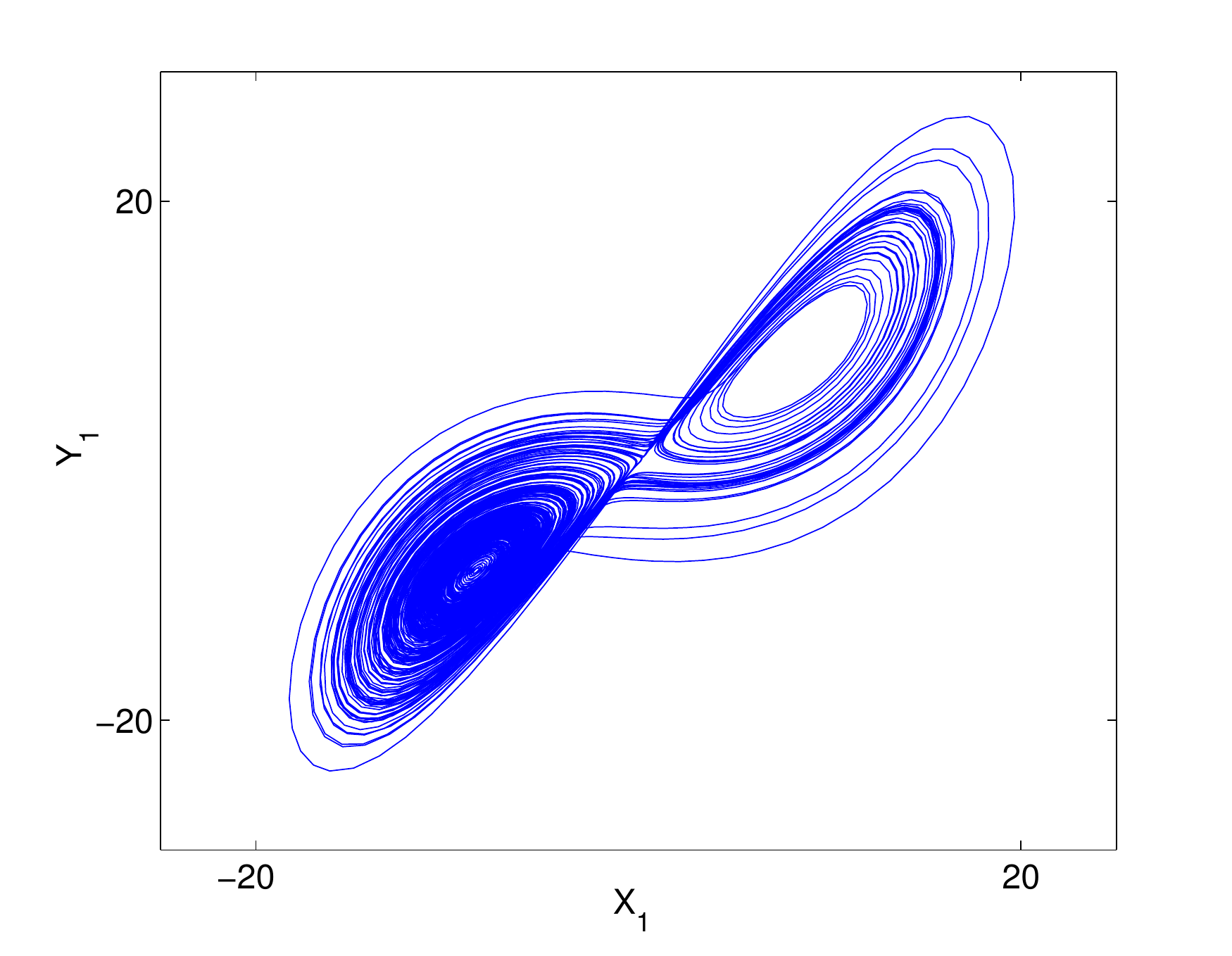}
			\caption{\label{pi}(a) Phase projection of $Y_1$ in System~(\ref{raodonghou}) with $Y_2$=1, $o_1$=1.5, $o_2$=0.2,
				$o_3$=1, $o_4$=2, $\varsigma$=1, $\sigma$=0.5 and $\tau$=10.}
		\end{figure}
		
		\begin{figure}
			\includegraphics[width=3in]{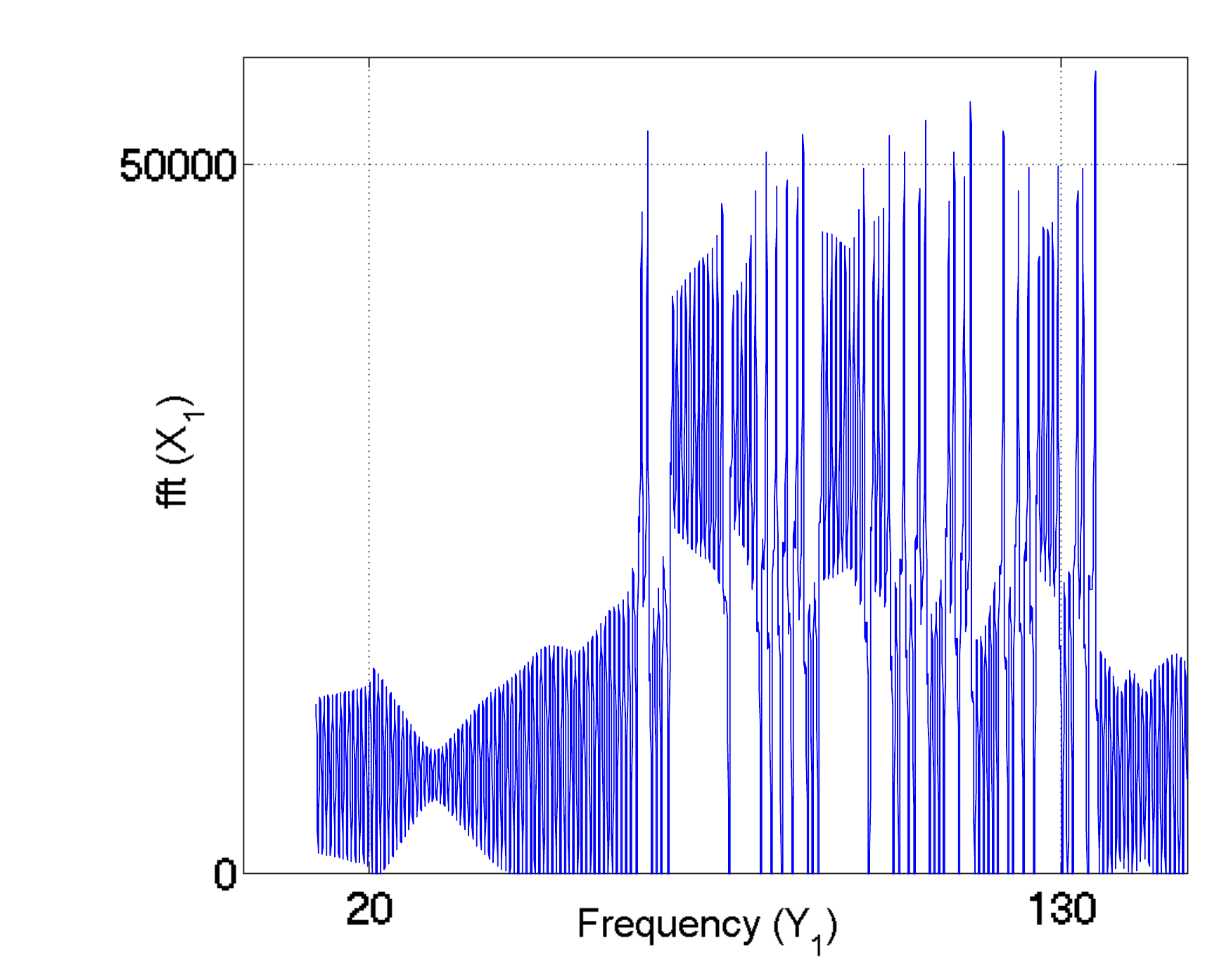}
			\caption{\label{pi}(b) Power spectra for $Y_1$ in System~(\ref{raodonghou}) which correspond with Fig.~3(a).}
		\end{figure}
		
		\begin{figure}
			\includegraphics[width=3in]{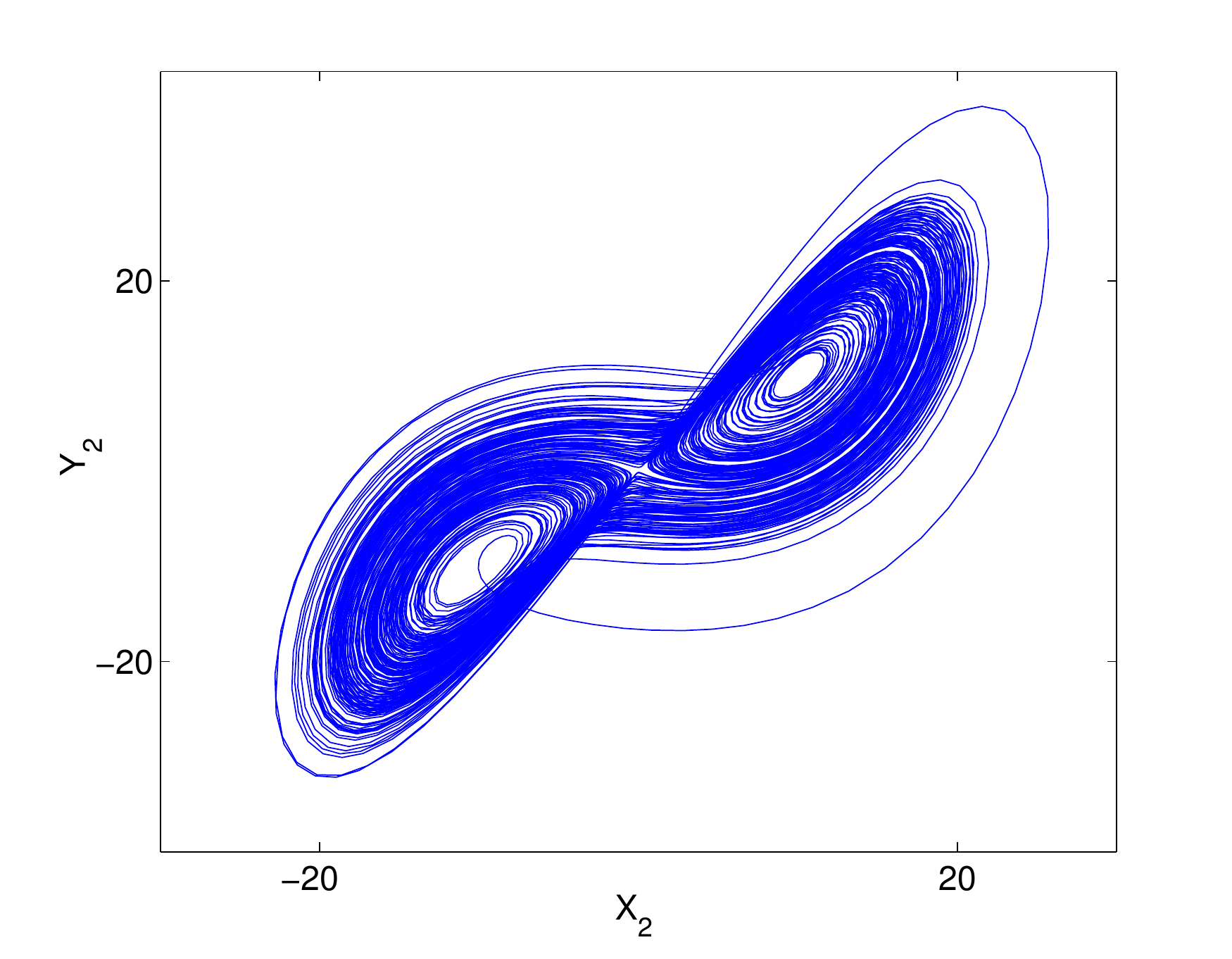}
			\caption{\label{pi}(a) Phase projection of $Y_2$ in System~(\ref{raodonghou}) with $Y_1$=1, $o_1$=1.5, $o_2$=0.2,
				$o_3$=1, $o_4$=2, $\varsigma$=1, $\sigma$=0.5 and $\tau$=10.}
		\end{figure}
		
		\begin{figure}
			\includegraphics[width=3in]{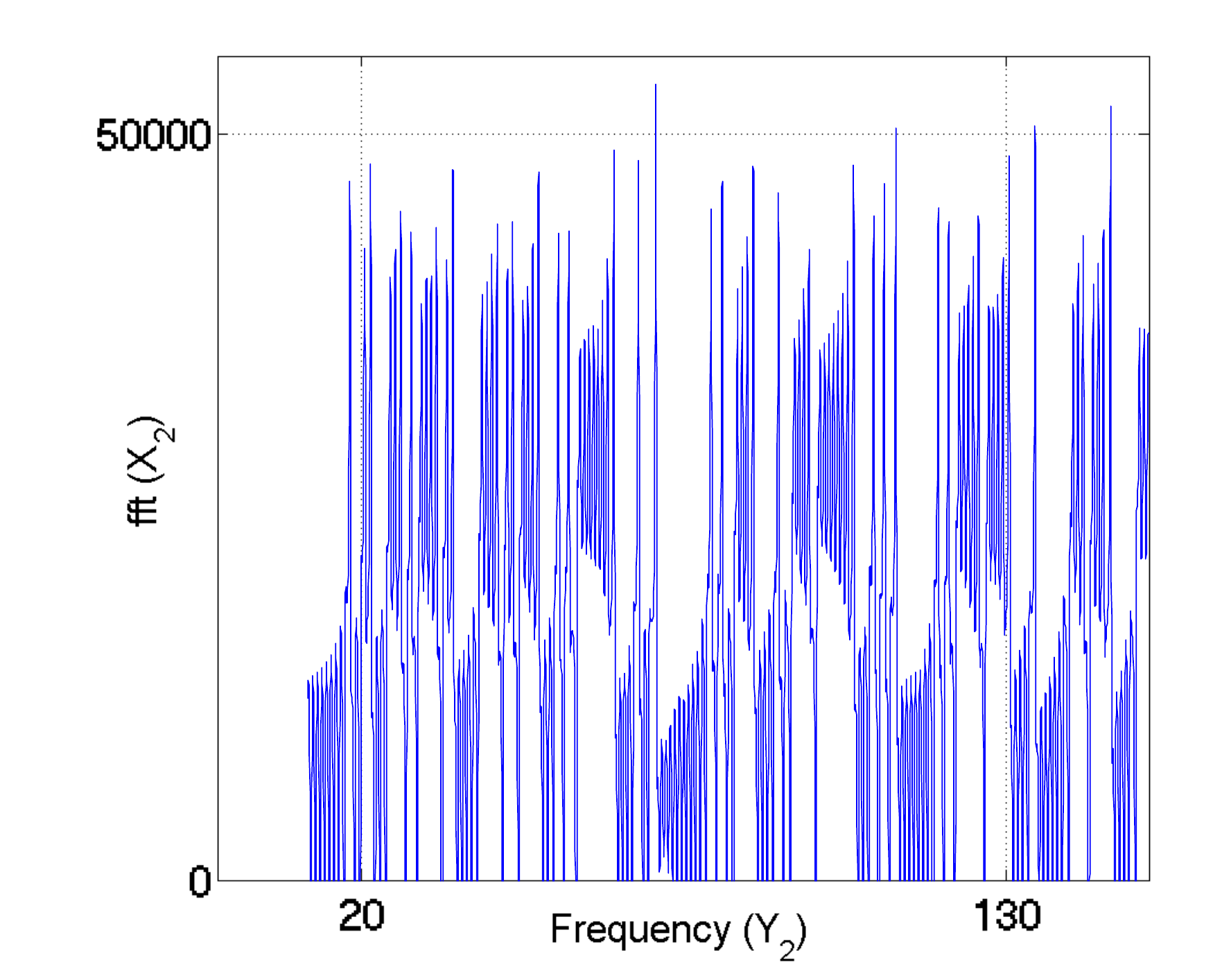}
			\caption{\label{pi}(b) Power spectra for $Y_2$ in System~(\ref{raodonghou}) which correspond with Fig.~4(a).}
		\end{figure}
		
		To study the final chaotic states of System~(\ref{raodonghou}), we investigate the
		phase projections of $Y_1$ and $Y_2$ in Figs.~3(a) and~4(a), respectively, and calculate
		their respective power spectra in Figs.~3(b) and~4(b). Comparing Figs.~3(b) with~2(a),~4(b)
		with~2(b), respectively, we can see that the original frequencies have been both broken, and
		chaotic motions occur. Note that the solutions of System~(\ref{raodonghou}) ignore
		the driver periods and represent a random sequence of uncorrelated shocks, so those
		chaotic motions are the ``developed" ones~\cite{developed1,developed2}.
		
		In this paper, we have discussed the CNLS equations [i.e., Eqs.~(\ref{fangcheng1})],
		which describe the propagation of envelopes of electromagnetic waves in birefringent
		optical fibers, for their potential applications in the secured optical communication.
		With the time-delay and perturbations introduced into Eqs.~(\ref{fangcheng1}), we
		have constructed a chaotic system and its final chaotic motions, with the phase
		projections and power spectra given. Further, soliton solutions and soliton propagation
		of such chaotic system have been studied when time-delay is fixed. As a generalization,
		the main results of this paper can be summarized as follows:
		
		$\bullet$ Reducing Eqs.~(\ref{fangcheng1}) into the equivalent four-dimensional planar
		dynamic system [i.e., System~(\ref{xitong})], we have obtained the integrability and
		periodicity of Eqs.~(\ref{fangcheng1}) from the phase projections and power spectra,
		as displayed in Figs.~1-2.
		
		$\bullet$ With time-delay and perturbations into System~(\ref{xitong}), we have chaotified
		Eqs.~(\ref{fangcheng1}) and a chaotic system [i.e., System~{\ref{raodonghou}}] has been
		constructed.
		
		$\bullet$ Chaotic motions of System~(\ref{raodonghou}) have been displayed via the phase
		projections, as shown in Figs.~3(a) and~4(a), and the respective power spectra have been
		calculated in Figs.3(b) and~4(b).
		
		\noindent\textbf {Acknowledgments}
		The authors acknowledge *** for the discussions during the works.

	\end{CJK*}
\end{document}